\documentclass[aps,prb,twocolumn,superscriptaddress]{revtex4-2}
\usepackage{xcolor}
\usepackage{amsmath,amssymb,mathtools}
\usepackage{commath}
\usepackage{bm,bbm}
\usepackage{physics}
\usepackage{tikz}
\usepackage{CJKutf8}
\usepackage[colorlinks = true,
            linkcolor = blue,
            urlcolor  = blue,
            citecolor = blue,
            anchorcolor = blue]{hyperref}
\usepackage{cleveref}
\usepackage{graphicx}
\usepackage{xcolor}
\usepackage{epstopdf}
\begin{document}
\crefname{equation}{Eq.}{Eqs.}
\crefname{figure}{Fig.}{Fig.}
\crefname{appendix}{Appendix}{Appendix}
\DeclarePairedDelimiter{\floor}{\lfloor}{\rfloor}
\newcommand{\proj}[1]{|{#1}\rangle\!\langle{#1}|}
\newcommand{\id}{\mathbbm1}
\newcommand{\G}{\mathcal G}
\newcommand{\D}{\mathcal D}

\definecolor{purple}{rgb}{.6,.1,.6}
\definecolor{darkgreen}{rgb}{.1,.6,.1}
\newcommand{\ZYW}[1]{\textcolor{purple}{ZYW:#1}}
\newcommand{\ZYWe}[1]{\textcolor{purple}{#1}}
\newcommand{\EC}[1]{\textcolor{blue}{[EC: #1]}}
\newcommand{\dm}[1]{\textcolor{darkgreen}{DM: #1}}
\newcommand{\JN}[1]{\textcolor{red}{#1}}
\newcommand{\jr}[1]{\textcolor{olive}{#1}}


\title{Measurement-induced entanglement in noisy 2D random circuits}
\date{\today}

\author{Zhi-Yuan Wei \begin{CJK}{UTF8}{gbsn}(魏志远)\end{CJK}}
\email{zywei@umd.edu}
\affiliation{Joint Quantum Institute and Joint Center for Quantum Information and Computer Science,
NIST/University of Maryland, College Park, Maryland 20742, USA}
\author{Jon Nelson}
\affiliation{Joint Quantum Institute and Joint Center for Quantum Information and Computer Science,
NIST/University of Maryland, College Park, Maryland 20742, USA}

\author{Joel Rajakumar}
\affiliation{Joint Quantum Institute and Joint Center for Quantum Information and Computer Science,
NIST/University of Maryland, College Park, Maryland 20742, USA}

\author{Esther Cruz}
\affiliation{
Max-Planck-Institut f{\"{u}}r Quantenoptik, Hans-Kopfermann-Str. 1, 85748 Garching, Germany
}
\affiliation{
Munich Center for Quantum Science and Technology (MCQST), Schellingstr. 4, 80799 M{\"{u}}nchen, Germany
}

\author{Alexey V. Gorshkov}
\affiliation{Joint Quantum Institute and Joint Center for Quantum Information and Computer Science,
NIST/University of Maryland, College Park, Maryland 20742, USA}

\author{Michael J. Gullans}
\affiliation{Joint Quantum Institute and Joint Center for Quantum Information and Computer Science,
NIST/University of Maryland, College Park, Maryland 20742, USA}
\affiliation{National Institute of Standards and Technology, Gaithersburg, MD 20899, USA.}

\author{Daniel Malz}
\affiliation{Department of Mathematical Sciences, University of Copenhagen, 2100 Copenhagen, Denmark}

\begin{abstract}
We study measurement-induced entanglement (MIE) generated by column-by-column sampling of noisy 2D random circuits of size $N$ and depth $T$. Focusing primarily on Clifford circuits and using the operator entanglement $S_{\rm op}$ of the sampling-induced boundary state as a proxy for computational complexity, first, we reproduce in the noiseless limit a finite-depth transition from area- to volume-law scaling at a threshold depth $T_c=6$. In contrast, in the presence of single-qubit depolarizing noise at any constant rate $p>0$, we find that the operator entanglement $S_{\rm op}$ obeys an area law, with its maximum value scaling approximately linearly with $T/p$ in the regime $T>T_c$. By analyzing the spatial distribution of stabilizer generators, we observe exponential localization of stabilizer generators; this both accounts for the scaling of the maximal $S_{\rm op}$ and implies an exponential decay of conditional mutual information across buffered tripartitions, which we also confirm numerically. Together, these results indicate that constant local noise destroys long-range MIE in 2D random Clifford circuits, and that a tensor-network–based algorithm can efficiently sample from noisy 2D random Clifford circuits (i) at sub-logarithmic depths $T = o(\log N)$ for any constant noise rate $p = \Omega(1)$, and (ii) at constant depths $T = O(1)$ for noise rates $p = \Omega(\log^{-1}N)$. Finally, we turn to depth $T=4$ Haar-random and measurement-based quantum computing-type circuits, providing evidence that MIE in noisy 2D Haar-random circuits exhibits the same qualitative behavior as in random Clifford circuits, and that noise destroys the volume-law scaling of MIE in non-Clifford circuits.

\end{abstract}

\maketitle
\section{Introduction}
\label{sec1}

Entanglement is widely recognized as a key resource enabling tasks unattainable in a classical world, including unconditionally secure cryptography and computational speed-ups. Indeed, generic highly entangled many-body states are believed to be hard to describe and simulate on classical hardware, and creating a large amount of entanglement is therefore a natural ingredient in any bid for quantum advantage. In unitary-only architectures constrained by spatial locality, generating such complexity typically demands deep circuits: for instance, states prepared by logarithmic-depth 1D circuits or constant-depth 2D circuits admit efficient classical algorithms for estimating local observables~\cite{Bravyi2021}. Somewhat counterintuitively, however, \emph{sampling} from constant-depth 2D circuits is worst-case hard~\cite{terhal2004adaptivequantumcomputationconstant} and is widely conjectured to be average-case hard~\cite{hangleiter2023computational}. Sampling can be hard because local measurements can convert short-range entanglement into long-range entanglement. This is well understood in the context of entanglement swapping~\cite{PhysRevLett.71.4287,PhysRevLett.80.3891,PhysRevA.57.822}, measurement-based quantum computation~\cite{Raussendorf2001,briegel2009measurement}, and efficient preparation of topologically ordered states~\cite{Raussendorf2005,Piroli2021a,lu2022measurement,iqbal2024non,PhysRevLett.131.200201,PhysRevX.14.021040}.
In the case of random circuit sampling, sampling from the bulk of the qudits in a state prepared by a constant-depth random circuit can yield a highly entangled circuit on the remaining qudits, a phenomenon called measurement-induced entanglement (MIE)~\cite{Hoke:2023aa}.

Despite the importance of understanding MIE in 2D circuits and its relation to simulation complexity, most prior studies have concentrated on the \emph{noiseless} setting~\cite{napp2022,PhysRevB.106.144311,PhysRevLett.132.030401,PRXQuantum.6.010356,PhysRevX.15.021059} (with the exception of Ref.~\cite{Hoke:2023aa}, which experimentally studied MIE in a noisy quantum processor). In realistic devices, however, noise is unavoidable. As noise suppresses entanglement growth in 1D random-circuit dynamics~\cite{Noh2020,li2023entanglement}, it is natural and important to ask whether the volume-law scaling of MIE in 2D random circuits can survive in the presence of noise. This question in 2D is conceptually significant and directly relevant to the classical simulability of \emph{noisy} 2D random-circuit sampling~\cite{hangleiter2023computational}, as the MIE serves as an indicator of the efficiency of the ``space-evolving block decimation'' (SEBD) algorithm~\cite{napp2022} (and its extension in \cref{sec_setup}) for sampling from 2D random circuits. 

It is known that, when  circuit depth scales at least logarithmically with system size, any depolarizing noise of constant rate $p>0$ renders 2D random-circuit sampling amenable to efficient classical simulation~\cite{10.1145/3564246.3585234,schuster2024polynomial, nelson2025limitationsnoisygeometricallylocal}. By contrast, in the \emph{sub-logarithmic} depth regime, efficient classical simulation under depolarizing noise has been established only at large noise rates $p>p_c$, via approaches such as trajectory unraveling~\cite{PRXQuantum.4.040326} and percolation-based analyses~\cite{nelson2024polynomialtimeclassicalsimulationnoisy,doi:10.1137/1.9781611978322.30}. Clarifying the fate of MIE at small, constant noise rate $p>0$ and at sub-logarithmic depths would not only illuminate the interplay among unitary gates, measurements, and noise, but also provide crucial insight into closing the remaining gap in our understanding of the simulation complexity of noisy 2D random-circuit sampling.

\begin{figure*}
	\centering
	\includegraphics[width=1\textwidth]{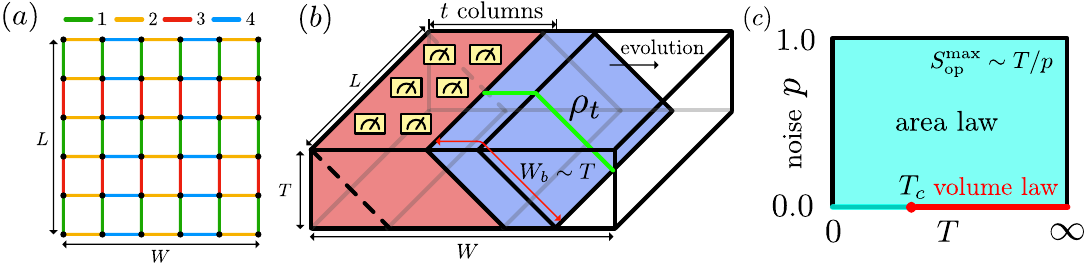}
        \caption{\emph{Setup and conjectured phase diagram.}
        \textbf{(a)} Brickwork layer order. Edges of the same color denote two-qubit gates applied within a single layer; the four-layer cycle (green $\rightarrow$ yellow $\rightarrow$ red $\rightarrow$ blue) is repeated periodically to entangle the lattice. One cycle corresponds to a circuit depth of $4$.
        \textbf{(b)} Column-by-column sampling process~\cite{napp2022}.  After every circuit layer, an on-site noise channel of rate $p$ acts independently on each qubit. Sampling proceeds along the width direction. To sample the $t$-th column ($t=3$ illustrated here), we include the gates and qubits within its past light cone (blue region) in the boundary state $\rho_t$ during the sampling procedure, and then measure the $t$-th column. The boundary state on the unmeasured region has length $L$ and width $W_b$~\cite{vftn_fn}.
We diagnose its half-space operator entanglement $S_{\rm op}$ across the cut along the direction of the width (green line), and also study conditional mutual information (CMI) in~\cref{sec_clif_string}. \textbf{(c)} Conjectured entanglement phase diagram of the maximal measurement-induced operator entanglement $S_{\rm op}^{\max}$, plotted versus noise rate $p$ and depth $T$. For $p=0$ (noiseless), a finite-depth transition at $T_c$ [cf.~\cref{fig2}(b)] separates volume-law and area-law behavior~\cite{napp2022}. For any constant noise rate $p>0$, our numerical and analytical analysis indicate area-law scaling of $S_{\rm op}^{\max}\sim T/p$ in the regime $T>T_c$.}
        \label{fig1}
\end{figure*}

In this work, we take a step towards this goal and investigate entanglement dynamics induced by column-by-column sampling of noisy 2D random \emph{Clifford} circuits. In contrast to general random circuits, these are efficiently simulatable~\cite{gottesman1998heisenberg,aaronson2004improved}, which allows us to numerically study their entanglement dynamics in large systems and at small noise rates. Since random Clifford circuits form a unitary 3-design~\cite{PhysRevA.96.062336} and often qualitatively reproduce entanglement dynamics of Haar-random circuits~\cite{PhysRevB.100.134306,rqc_review,PhysRevLett.132.030401}, we expect this Clifford-based simulation to capture the key features of entanglement during  sampling dynamics of generic noisy 2D random circuits. To quantify MIE, we simulate sampling the 2D state column-by-column and consider the operator entanglement $S_{\rm op}(t)$ of the boundary state $\rho_t$ generated by the sampling process (see the setup in \cref{sec_setup}). In \cref{sec_clif_num}, we numerically characterize the scaling of $S_{\rm op}$ in two regimes: in the noiseless limit $p=0$, we identify a finite-depth transition from area- to volume-law behavior with a critical depth $T_c= 6$, consistent with previous studies~\cite{napp2022,PhysRevX.15.021059,PRXQuantum.6.010356}. This transition is also conceptually related to the measurement-induced phase transition in random circuits~\cite{Skinner2019,PhysRevX.10.041020,PhysRevB.100.134306,PhysRevB.106.214316}. In stark contrast, for any constant noise rate $p>0$, the maximal $S_{\rm op}$ obeys an area-law (i.e., it is independent of the boundary length); in the regime $T>T_c$, it grows approximately linearly with depth, $S_{\mathrm{op}}^{\max}\sim \beta_p T$, with $\beta_p\sim p^{-0.91}$ decreasing as $p$ increases. In \cref{sec_clif_string}, we relate this scaling to the spatial distribution of stabilizer generators: stabilizer generators are distributed almost evenly in the bulk of the boundary state, while their length $\ell$ occur with probability ${\cal D}(\ell)\sim e^{-\gamma_{p,T}^{\rm len}\ell}$. This bounds the number of stabilizer generators crossing a bipartition, underpins the area-law scaling of $S_{\rm op}^{\rm max}$, and implies exponential decay of the conditional mutual information (CMI) of $\rho_{t_{\rm peak}}$—a behavior we also confirm numerically. We then show in \cref{sec_clif_Sop} that, assuming the numerically observed scalings hold, the MPO-SEBD algorithm [extended from the ``space-evolving block decimation''~(SEBD) algorithm~\cite{napp2022}, cf.~\cref{sebd_setup}] \emph{guarantees} efficient sampling from noisy 2D random Clifford circuits (i) throughout the entire sub-logarithmic depth regime $T = o(\log N)$ for arbitrarily small constant noise rate $p = \Omega(1)$, and (ii) at constant depths $T = O(1)$ for noise rates $p = \Omega(\log^{-1}N)$, within polynomial time.
In \cref{sec_haar}, we demonstrate the MPO-SEBD algorithm on the sampling dynamics of noisy 2D Haar-random circuits and measurement-based quantum computing (MBQC)-type circuits of depth $T=4$, providing evidence that the MIE in noisy 2D Haar-random and random Clifford circuits exhibits the same qualitative behavior, and that noise can destroy the volume-law scaling of MIE in non-Clifford circuits. Finally, in \cref{sec_implic}, we discuss the implications of our results for other classical simulation algorithms of noisy 2D circuits, and highlight how our results provide evidence that MIE in noisy, shallow 2D Haar-random circuits may also exhibit area-law scaling.

\section{Setup}
\label{sec_setup}
\subsection{Noisy 2D circuit sampling and the MPO-SEBD algorithm}
\label{sebd_setup}
We consider a geometrically local brickwork circuit of depth $T$ acting on a two-dimensional lattice of size $N=L\times W$, with qubits initialized in the product state $\ket{0}^{\otimes N}$ [cf.~\cref{fig1}(a)].
The sampling algorithm we consider here builds on the SEBD algorithm first introduced in Ref.~\cite{napp2022}.
In this algorithm, sampling proceeds column-by-column (we index the columns by $t$ from left to right) along the width direction [\cref{fig1}(b)], which has a desirable property that, in every step, we only need to keep track of the boundary state. 
More concretely, the algorithm proceeds as follows~\cite{napp2022}:
\begin{enumerate}
    \item To sample the first column ($t=1$), we consider only the gates contained in its past light-cone and the qubits they act on (the triangular-prism region).
    This defines a state $\rho_1$ supported on the Hilbert space corresponding to a strip of $L\times W_b$ qubits, with the width of the boundary state $W_b\sim T$~\cite{vftn_fn}.
    We then sample the first column according to the marginal of $\rho_1$, obtaining the bitstring $\vec b_1$ and the conditional state $\rho_{1}^\mathrm{cond}\propto\tr_1[(\proj{\vec b_1}\otimes\id)\rho_1]$ supported on $L\times (W_b-1)$ qubits.
    \item To sample the $t$-th column, we take the normalized conditional state from the previous step $\rho_{t-1}^\mathrm{cond}$ and add any qubits and gates in the past lightcone of the $t$-th column that were not included in $\rho_{t-1}$. This defines $\rho_t$. The marginal of the $t$-th column of that state is now the correct marginal of the state conditioned on the outcomes of the preceding columns. We sample from this marginal and proceed to the next column.
    \item Iterating this unitary–measure cycle yields a full sample from the output distribution of the constant-depth 2D circuit.
\end{enumerate}
Note that step 2 above describes the action of a POVM, and thus we can think of the sequence of states $\rho_t$ generated in the above sampling protocol as the time evolution of the boundary state $\rho_t$ under unitary evolution, measurement, and the addition of fresh qubits.

In the noiseless case, the boundary state remains pure, i.e., $\rho_t=\proj{\psi_t}$, and SEBD can be implemented by representing $\ket{\psi_t}$ as a matrix-product state~\cite{napp2022}.
Here, we consider a noisy version of the above protocol, in which each gate of the circuit is followed by single-qubit depolarizing noise at rate $p$.
As a result, the boundary state is no longer pure, and we can instead represent it as a matrix-product operator (MPO).
We refer to the resulting sampling protocol as the MPO-SEBD algorithm.
Note that this differs from the noisy-SEBD method of Ref.~\cite{PRXQuantum.4.040326}, which treats on-site noise via quantum trajectories, with each trajectory simulated by SEBD.

Note that the boundary state $\rho_t$ [cf.~\cref{fig1}(b)] spans a quasi-one-dimensional region containing $N_b=L\times W_b$ qubits. Representing $\rho_t$ as a one-dimensional MPO with $L$ sites then requires grouping the $W_b\sim T$ qubits along the width direction into a single qudit of dimension $d=2^{W_b}\sim 2^T$. Consequently, both the memory cost and the runtime of the MPO-SEBD algorithm grow exponentially with the circuit depth $T$.

\subsection{Measurement-induced operator entanglement}
\label{sec_Sop}
The key question we want to address is how complex the boundary state during the sampling process is.
The primary diagnostic we use for this is the \emph{operator entanglement} $S_{\rm op}$~\cite{PhysRevA.63.040304} across a fixed horizontal half-space bipartition [illustrated as the green line in \cref{fig1}(b)]. We define the vectorization of $\rho_t$ as $\ket{\rho_t}$, and the associated normalized density matrix for $\ket{\rho_t}$ as
\begin{equation}
\tilde\rho_t = \frac{\proj{\rho_t}}{\operatorname{tr}[\rho_t^2]}.
\end{equation}
For a bipartition $AB$, the operator entanglement is the von Neumann entropy of the reduced state of $\tilde\rho_t$~\cite{PhysRevA.63.040304},
\begin{equation}\label{eq:Sop_def}
S_{\rm op}(t) \coloneqq -\operatorname{tr}\big[\tilde\rho_t^{A}\log \tilde\rho_t^{A}\big] ,
\end{equation}
with $\tilde\rho_t^{A}= \tilde {\operatorname{tr}}_B [\tilde\rho_t]$, where $\tilde {\operatorname{tr}}$ denotes the traces of density matrices on the vectorized (doubled) Hilbert space.

\begin{figure*}
	\centering
	\includegraphics[width=\textwidth]{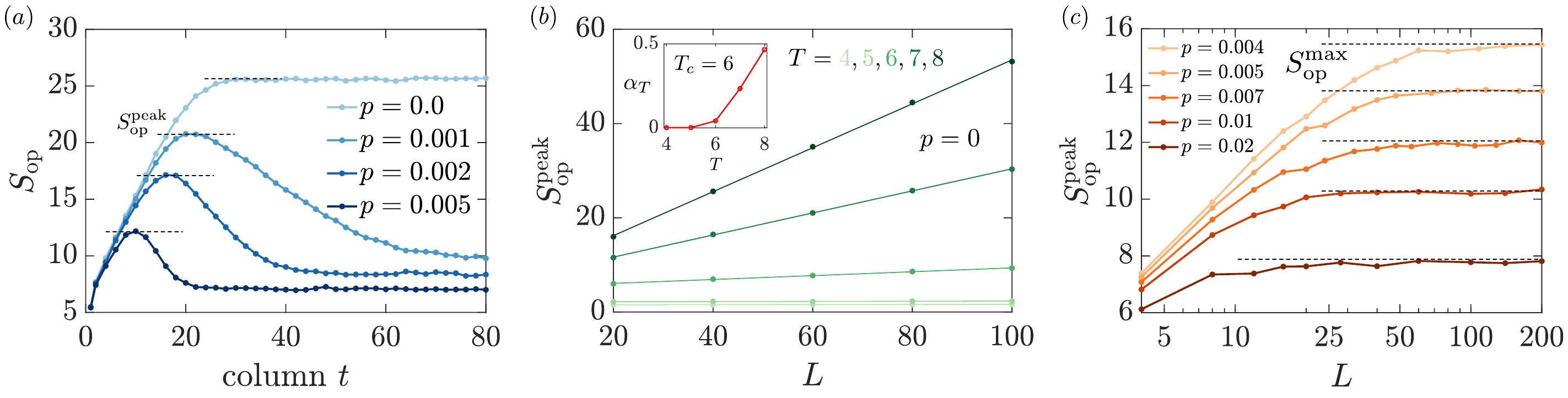}
        \caption{\emph{Dynamics and scaling of measurement-induced operator entanglement.} \textbf{(a)} Evolution of $S_{\rm op}(t)$ during sampling for noiseless and noisy circuits, with circuit depth $T=8$ and length $L=40$. For the noiseless case ($p=0$), $S_{\rm op}(t)$ increases monotonically with $t$ and saturates to its volume-law value. For noisy case ($p>0$), $S_{\rm op}(t)$ grows, reaches a peak value $S_{\rm op}^{\rm peak}$, and then decreases toward a steady-state value. \textbf{(b)} Finite-depth transition in the noiseless case. Plotted is $S_{\rm op}^{\rm peak}$ versus $L$ for depths $T\in4,5,6,7,8$ with linear fits $S_{\rm op}^{\rm peak}=\alpha_T L+c_T$. The inset shows the fitted slope $\alpha_T$ as a function of $T$, identifying a critical depth $T_c= 6$ where $\alpha_T$ turns nonzero. \textbf{(c)} Area-law scaling in the noisy case at fixed depth $T=8$. For each noise rate $p$, $S_{\rm op}^{\rm peak}$ exhibits a nearly logarithmic growth with $L$ at small $L$ and then saturates to the asymptotic maximal value $S_{\rm op}^{\rm max}$, indicated by the dashed lines.}
        \label{fig2}
\end{figure*}

In the noiseless case, $\rho_t$ is pure, and $S_{\rm op}$ is twice the von Neumann entropy of $\ket{\psi_b}$, directly tying $S_{\rm op}$ to the bond dimension required by matrix-product state simulations.
For mixed states, $S_{\rm op}$ plays an analogous role for MPO representations and has been used to characterize simulation cost in open system evolution~\cite{RevModPhys.93.015008}, noisy 1D circuits~\cite{Noh2020}, and noisy Gaussian boson sampling~\cite{PhysRevA.104.022407}. An important caveat is that truncating small singular values from an MPO only guarantees a small error in Frobenius ($L_2$) norm, and not in total-variation distance~\cite{PRXQuantum.1.010304}. Thus, an area-law scaling of the boundary-state operator entanglement does not, by itself, guarantee an efficient MPO representation of the corresponding density matrix. By contrast, a volume-law scaling of $S_{\rm op}$ rules out any efficient MPO representation. Moreover, in \cref{sec_clif_Sop} we show that, for noisy stabilizer states, an area-law scaling of $S_{\rm op}$ actually guarantees an efficient MPO representation.

Operationally, we evaluate $S_{\rm op}$ of the pre-measurement boundary states $\rho_t$.
The difference between this convention and computing $S_\mathrm{op}$ after the measurement corresponds to only an $O(1)$ additive shift and thus does not affect the asymptotic scaling of $S_{\rm op}$ with system size or circuit depth.

In addition to using $S_{\rm op}$ as our primary diagnostic, in \cref{sec_clif_string}, we also compute the conditional mutual information (CMI). For a tripartite system with regions $A$ and $B$ separated by $C$, the CMI is
\begin{equation}
	I(A:B|C)=I(A:BC)-I(A:C),
	\label{cmi_def}
\end{equation}
where the mutual information is $I(X:Y)=S_X+S_Y-S_{XY}$, and $S_X=-\operatorname{tr}[\rho_X\log\rho_X]$ denotes the von Neumann entropy. Note that our calculation of $S_{\rm op}$ and the CMI also implicitly conditions on the measurement outcomes in the previously sampled columns (see \cite{Anshu_2019} Fact 34.D for a simple proof), since we consider the boundary state after projective measurements.

\subsection{Noisy 2D random Clifford circuits and mixed stabilizer states}
\label{sec_clif_setup}
For our numerics, we use the setup in \cref{sebd_setup} with random two-qubit Clifford gates.
We simulate the depolarizing noise in a Monte-Carlo style by replacing any given qubit by the maximally mixed state with probability $p$. This ensures that the state remains a stabilizer state throughout the evolution, enabling efficient classical simulation~\cite{aaronson2004improved}.

A (mixed Pauli) stabilizer state on a chain of $N$ qubits (labelled from 1 to $N$) can be defined in terms of its set of generators $\G=\{g_1,\cdots,g_M\}$, which comprises $M\leq N$ mutually commuting Pauli strings, as
\begin{equation}
	\rho_\G\coloneqq\prod_{k=1}^M\frac{\id+g_k}{2}.
	\label{eq:stabilizer-state}
\end{equation}
If $N=M$, the state is pure.
For to each generator, we can associate a start and end, which we define as the first and last qubits on which the generator is nontrivial (not the identity).
A convenient way to write down $\G$ is the \emph{clipped gauge}, which is a unique form in which at most two generators start and end on any given site~\cite{Nahum2017,PhysRevB.100.134306}.
The reason this gauge is so convenient, is that we can directly read off entanglement properties of the state from its generator.

Specifically, consider a bipartition of the chain into a contiguous left chunk $A=[i]_1^{N_A}$ and right chunk $B=[i]_{N_A+1}^N$.
We split up the generator set into three disjoint sets $\{\G_A,\G_B, \G_{AB}\}$, where $\G_A$ and $\G_B$ contain all generators wholly contained in either $A$ and $B$, and $\G_{AB}$ contains all generators with support in both $A$ and $B$.
Since the generators commute, we can write $\rho_\G$ as 
\begin{equation}
	\begin{aligned}
	\rho_\G
	&= (\rho_{\G_A}\otimes\rho_{\G_B})\rho_{\G_{AB}}\\
	&=
	(\rho_{\G_A}\otimes\rho_{\G_B})\sum_{\vec x\in\{0,1\}^{M_{AB}}}\prod_{g_i\in \G_{AB}}\frac{g_i^{x_i}}{2},
	\end{aligned}
	\label{eq:G-AB}
\end{equation}
where $\rho_{\G_{A/B}}$ are the reduced states on $A$ and $B$, which equal the stabilizer state defined through $\G_{A/B}$.
\Cref{eq:G-AB} is a sum over $2^{M_{AB}}$ terms, where $M_{AB}$ is the number of generators that have support in both $A$ and $B$.
This implies that the bond dimension $\chi_{AB}$ required to represent this state as MPO is $2^{M_{AB}}$.
From \cref{eq:G-AB}, we can also evaluate the operator entanglement \cref{eq:Sop_def}, which evaluates to $M_{AB}$.
The entropy of a stabilizer state of the form \cref{eq:stabilizer-state} is equal to $N-M$. Thus, we can directly read off the mutual information, which also equals the number of generators with support in both $A$ and $B$.
Thus, we have 
\begin{equation}
    S_\mathrm{op} = I(A : B) = \log_2(\chi_{AB}) = M_{AB}.
    \label{eq:string_ent}
\end{equation}
The CMI \cref{cmi_def}, which is defined in terms of a tripartition $A,B,C$, where $C$ buffers $A$ and $B$ similarly evaluates to the number of generators that have support both in $A$ and $B$ (but note that now the definition of $A$ and $B$ is different).

\section{Numerical results}
\label{sec_clif_num}

In this section, we present numerics for random Clifford circuits based on the stabilizer formalism, where each data point is averaged over $\sim 10^3$ individual realizations of the random circuits. These results, together with prior noiseless results~\cite{napp2022}, motivate the conjectured entanglement phase diagram in \cref{fig1}(c). There, for any constant noise rate $p>0$, the maximally attainable operator entanglement [denoted as $S_{\rm op}^{\max}$ (cf.~\cref{smax_def})] during sampling of noisy 2D random Clifford circuits exhibits area-law scaling.

The typical evolution of $S_{\rm op}(t)$ for the boundary state during the sampling process [cf.~\cref{fig1}(b)] is shown in \cref{fig2}(a). In the noiseless case $p=0$, $S_{\rm op}$ grows monotonically with the column index $t$ and saturates at a volume-law value. With noise $p>0$, $S_{\rm op}$ increases during the early iterations of the algorithm, reaches a peak $S_{\rm op}^{\rm peak}$ when sampling the $t_{\rm peak}$-th column of the qubits, and then decreases toward a steady-state value. This behavior reflects the interplay between entangling unitary layers, noise that suppresses quantum correlations, and sampling measurements. Similar non-monotonic profiles have been reported in 1D and noisy 2D circuit dynamics~\cite{Noh2020,zhang2022entanglement,li2023entanglement}.

In the following, we focus on the peak value $S_{\rm op}^{\rm peak}$, as it captures the maximal operator entanglement generated during the sampling process and indicates the largest bond dimension required to represent the boundary state $\rho_t$ as an MPO. We define
\begin{equation}
S_{\rm op}^{\rm peak}(L,T,p)=\max_{\textrm{column }t} S_{\rm op}(t).
\end{equation}

We first examine $S_{\rm op}^{\rm peak}$ versus the length $L$ of the boundary state in the noiseless case $p=0$, as shown in \cref{fig2}(b), where we observe a generic scaling
\begin{equation}
S_{\rm op}^{\rm peak}(L,T,0)\approx \alpha_T L+c_T,
\end{equation}
with the volume-law coefficient $\alpha_T$ becoming nonzero at a critical depth $T_c=6$ [inset of \cref{fig2}(b)]. This indicates an area-to-volume-law transition as $T$ increases~\cite{napp2022,PhysRevX.15.021059,PRXQuantum.6.010356}, and the extracted $T_c=6$ agrees with Ref.~\cite{PRXQuantum.6.010356}. Since noise can only reduce $S_{\rm op}$, this implies area-law scaling of $S_{\rm op}^{\rm peak}$ for all $T<T_c$ for arbitrary $p\ge 0$. In what follows we therefore focus on $T> T_c$.

\begin{figure}[h!]
	\centering
	\includegraphics[width=0.45\textwidth]{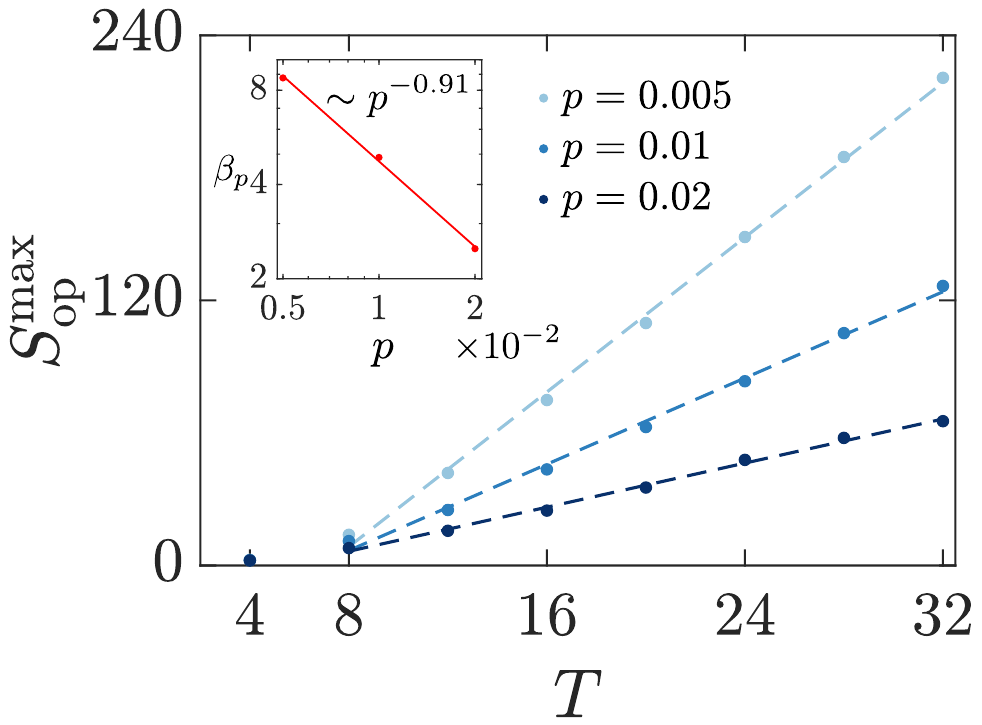}
        \caption{Saturated area-law value $S_{\rm op}^{\rm max}$ versus circuit depth $T$ for several fixed noise rates $p$. The dashed lines are linear fits $S_{\rm op}^{\rm max}(T,p)=\beta_p T+c_p$. Inset: fitted slopes $\beta_p$ plotted against $p$. The line is a power-law fit $\beta_p\sim p^{-0.91}$ to the data.}
        \label{fig3}
\end{figure}

Figure~\ref{fig2}(c) displays the scaling of $S_{\rm op}^{\rm peak}$ with $L$ at fixed depth $T=8$ and several noise rates $p>0$. For small $L$, we observe a slow, nearly logarithmic growth with $L$, followed by saturation to a plateau value $S_{\rm op}^{\rm max}$, defined as
\begin{equation}\label{smax_def}
S_{\rm op}^{\rm max}(T,p)=\lim_{L\to\infty} S_{\rm op}^{\rm peak}(L,T,p),
\end{equation}
which demonstrates an asymptotic area-law in the noisy case, in contrast to the volume-law at $p=0$ for the same depth. The scaling of the saturated value $S_{\rm op}^{\rm max}$ is summarized in \cref{fig3}, where we find
\begin{equation}\label{smax_scale}
S_{\rm op}^{\rm max}(T,p)\approx \beta_p T+c_{p},
\end{equation}
with a slope $\beta_p\sim p^{-0.91}$ that decreases with increasing noise.

\section{distribution of stabilizer generators}
\label{sec_clif_string}
\begin{figure*}
	\centering
	\includegraphics[width=\textwidth]{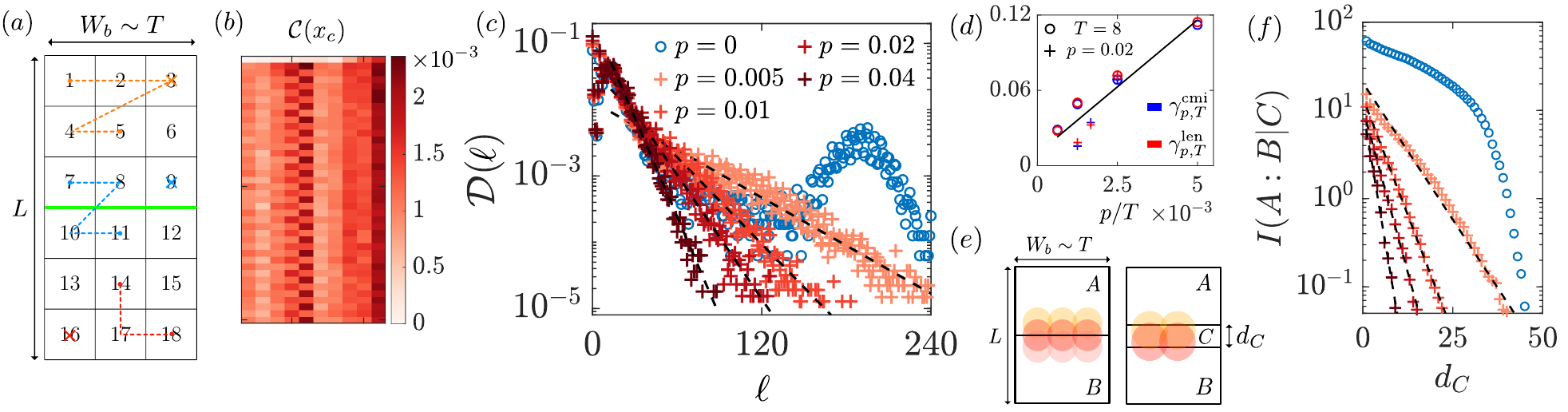}
        \caption{\emph{Distribution of stabilizer generators and the behavior of conditional mutual information (CMI) at $t_{\rm peak}$}. \textbf{(a)} 1D labeling of the qubit sites for the boundary state $\rho_{t_{\rm peak}}$ (here $L=6$, $W_b=3$). The green line denotes the horizontal half-space bipartition for computing $S_{\rm op}$ [cf.~\ \cref{fig1}(b)]. Colored dashed segments are examples of stabilizer generators, with circles marking their left endpoints $x_l$ and right endpoints $x_r$. The center locations of the stabilizer generators, com($g$)$=[x_l(g)+x_r(g)]/2$, are marked with crosses. \textbf{(b)} The spatial center distribution ${\mathcal C}(x_c)$ of stabilizer generators. Here $p=0.01,L=80,T=8$. The labeling of qubit sites follows the convention from panel (a). Since $\mathrm{com}(g)$ can take half-integer values, we place each half-integer site $x + 1/2$ immediately to the right of its corresponding integer site $x$. \textbf{(c)} Normalized length distribution $\mathcal D(\ell)$ of stabilizer generators versus $\ell$ for fixed depth $T=8$, system length $L=80$, and several noise rates $p$. The dashed lines are exponential fits to the data for noisy cases.
\textbf{(d)} The extracted exponential scaling exponent $\gamma_{p,T}^{\rm len}$ [cf.~\cref{clif_len_scale}] of the stabilizer generator length distribution, as well as that of the CMI, $\gamma_{p,T}^{\rm CMI}$. For fixed depth $T=8$ (circle markers), we varied the noise rate $p\in{0.005,0.01,0.02,0.04}$. For fixed $p=0.02$ (plus markers), we studied $T\in{8,12,16}$. The line shows a single linear fit to the combined data for $\gamma_{p,T}^{\rm len}$ and $\gamma_{p,T}^{\rm CMI}$. The plotted exponents for $T=8$ are extracted from the dashed lines in panels (c) and (f), while the corresponding plots of the length distribution and CMI for $T=12,16$ are omitted. \textbf{(e)} Schematic of the bipartition $AB$ and tripartition $ACB$ with buffer region $C$ of length $d_C$. For the bipartition, only the stabilizer generators (illustrated as colored circles) that cross the cut contribute to $S_{\rm op}$ [cf.~\cref{eq:string_ent}]. For the tripartition, only the stabilizer generators that cross the buffer region $C$ contribute to the CMI $I(A:B|C)$.
\textbf{(f)} The scaling of CMI $I(A:B|C)$ with the length $d_C$ of the buffer region $C$. The markers and the corresponding parameters ($T=8,L=80$) are the same as in panel (c). The dashed lines are exponential fits to the data for noisy cases.}
        \label{fig4}
\end{figure*}

To elucidate the origin of the area-law scaling of the maximal operator entanglement $S_{\rm op}^{\rm max}$ [cf.~\cref{smax_scale}] in the regime $T>T_c$, we analyze the spatial distribution of stabilizer generators in the clipped gauge of the boundary state $\rho_{t_{\rm peak}}$, at the time $t_{\rm peak}$ at which $S_{\rm op}$ attains its maximal value. We label boundary qubits by integers increasing along the width direction as in \cref{fig4}(a), 
bring the stabilizer tableau of $\rho_{t_{\rm peak}}$ to the clipped gauge~\cite{Nahum2017,PhysRevB.100.134306}, and obtain a set of stabilizer generators $\mathcal{G}=\{g_1,\dots,g_M\}$ acting nontrivially on subsets of sites. Here the number of stabilizer generators $M$ depends on the individual realization of the noisy circuit as well as the measurement outcome.
To each generator $g$ we can associate a left end $x_l(g)$ and a right end $x_r(g)$, which we define as the first and last qubits on which the generator is nontrivial (not the identity).
We also define the length and center of a stabilizer generator as
\begin{equation}
\begin{aligned}
    {\rm len}(g)&=x_r(g)-x_l(g), \\
    {\rm com}(g)&=[x_l(g)+x_r(g)]/2.
\end{aligned}
\end{equation}

To obtain the distributions of the centers and lengths of stabilizer generators, we produce a collection of boundary states $\{\rho_{t_{\rm peak}}\}$ across many independent numerical realizations of the sampling process, from which we gather a total of $N_g \sim 10^5$ stabilizer generators $\{g_i\}_{i=1}^{N_g}$ across all realizations.

We characterize the spatial structure of stabilizer generators via two distributions: the \textit{center-location distribution} $\mathcal{C}(x_c)$ and the \textit{length distribution} $\mathcal{D}(\ell)$. They are defined as the probabilities that a randomly picked stabilizer generator $g$ has center $\mathrm{com}(g)=x_c$ and length $\mathrm{len}(g)=\ell$, respectively. These distributions can be computed numerically as
\begin{equation}
\mathcal{C}(x_c)=\frac{1}{N_g}{\sum_{i=1}^{N_g}\delta_{\operatorname{com}(g_i),x_c}},
\end{equation}
\begin{equation}
\mathcal{D}(\ell)=\frac{1}{N_g}{\sum_{i=1}^{N_g}\delta_{\operatorname{len}(g_i),\ell}}.
\label{eq:D}
\end{equation}

The numerically obtained $\mathcal{C}(x_c)$ at $t_{\rm peak}$ is shown in \cref{fig4}(b) for depth $T=8$ and noise rate $p=0.01$ as a representative example. Within the bulk of the boundary state $\rho_{t_{\rm peak}}$ that consists of $N_b$ qubits, $\mathcal{C}(x_c)$ remains approximately uniform along both the length and width, i.e. $\mathcal{C}(x_c)\approx 1/N_b$. Here the uniformity is expected, since the unitary gates, noise, and measurements are applied nearly uniformly during the column-by-column sampling process [cf.~\cref{fig1}(b)].

\Cref{fig4}(c) shows the length distribution $\mathcal{D}(\ell)$ at $t_{\rm peak}$ for circuit depth $T=8$ in the noiseless case $p=0$ and noisy case with $p=0.01,0.02,0.04$. At $p=0$, where $S_{\rm op}^{\rm max}$ saturates to a volume-law value, a substantial fraction of long stabilizers is visible. If $p>0$, the distribution acquires an exponential decay for $\ell$ larger than certain threshold $\ell_0$ [$\ell_0 \approx10$ in \Cref{fig4}(c)], as
\begin{equation}\label{clif_len_scale}
\mathcal{D}(\ell>\ell_0)\propto e^{-\gamma_{p,T}^{\rm len}\ell},
\end{equation}
with an exponent $\gamma_{p,T}^{\rm len} \approx \eta p/T$, as shown in \cref{fig4}(d), with numerically extracted $\eta\approx 21.3$. The exponential decay indicates noise-induced localization of stabilizer generators in the clipped gauge.
Intuitively, this stems from the interplay of noise, unitaries, and measurements. Noise tends to remove stabilizer generators, whereas measurement introduces single-site stabilizer generators, and unitaries lengthen them. This picture is not wholly predictive, as measurements can lead to a complex reshuffling of stabilizer generators in the clipped gauge, but it suggests that exponential decay of long stabilizer generators is the generic behavior in these dynamics.

\subsection{From the stabilizer distribution to the observed entanglement scaling}
\label{sec_clif_distri_Sop}

We now show that the numerically observed exponential suppression of long stabilizer generators, $\mathcal D(\ell)\sim e^{-\gamma_{p,T}^{\rm len} \ell}$, explains the scaling in \cref{smax_scale}. We consider in particular the asymptotic value ($L\gg 1$) of the maximum operator entanglement as $S_\mathrm{op}^\mathrm{max}$.
The observed center-location and length distribution of stabilizer generators motivate us to consider the following simplified model, where the probability that a randomly picked stabilizer generator $g$, with an arbitrary \emph{left endpoint} $x_l$, has length at least $\ell$ is exponentially suppressed:
\begin{equation}
    P(\mathrm{len}(g)\ge \ell)=e^{-\gamma_{p,T}^{\rm len}\ell}.
    \label{eq:suppression}
\end{equation}
This distribution omits the small-size correction for $\ell <\ell_0$ and the small spatial non-uniformity of the spatial distribution of stabilizer generators, which do not alter the predicted scaling.

At $t_{\rm peak}$, where $S_{\rm op}$ reaches its maximum value, in the numerical regimes we studied, we observe that the average density of stabilizer generators satisfies $\langle M \rangle / N \approx 1$, since noise has not yet accumulated significantly within the finite time $t_{\rm peak}$. Then, using \cref{eq:string_ent}, the expected peak operator entanglement $S_\mathrm{op}^\mathrm{max}$ can be obtained by summing the probabilities of stabilizer generators $g$ whose left endpoint lies in region $A$ (i.e., $x_l \le N_b/2$) and whose right endpoint $x_r$ lies in region $B$ (i.e., $x_r > N_b/2$). Summing over all lattice sites in region $A$ then yields the following expression for $S_\mathrm{op}^{\rm max}$
\begin{align}\label{S_model_scale}
     S_\mathrm{op}^{\rm max}&\approx \sum_{\ell=1}^{N_b/2\rightarrow \infty} P(\mathrm{len}(g)\ge \ell) \\
    &= 1/(e^{\gamma_{p,T}^{\rm len}}-1)\approx 1/\gamma_{p,T}^{\rm len}
    \approx T/(\eta p) \nonumber
\end{align}
where we used the fact that $\gamma_{p,T}^{\rm len} \ll 1$ for $p\ll 1$ in the last step [cf.~\cref{fig4}(d)]. The predicted behavior, $S_\mathrm{op}^{\rm max} \approx  T/(\eta p)$, exhibits good quantitative agreement with the numerical observations presented in \cref{fig3}.

Similarly, we can use \cref{eq:suppression} to predict the scaling of conditional mutual information across a buffered tripartition $ACB$ with buffer width $d_C$ [see the setup illustrated in \cref{fig4}(e)]:
\begin{equation}\label{cmi_predict}
    I(A:B|C) \approx \sum_{\ell=W_bd_C+1}^\infty P(\mathrm{len}(g)\ge \ell)
    = \frac{e^{-\gamma_{p,T}^{\rm len} W_b d_C}}{e^{\gamma_{p,T}^{\rm len}}-1}.
\end{equation}
Hence the CMI is expected to decay exponentially with the buffer size $d_C$ with an exponent $\gamma_{p,T}^{\rm len} W_b$.
We verify this numerically in \cref{fig4}(f), with the tripartition $ACB$ as shown in \cref{fig4}(e). For the noiseless case $p=0$, we observe long-range CMI, as expected from previous studies~\cite{napp2022,PhysRevB.106.144311,PhysRevLett.132.030401,PRXQuantum.6.010356,PhysRevX.15.021059}. For noisy cases $p>0$, we see exponential decay of CMI $I(A:B|C) \sim  e^{-\gamma_{p,T}^{\rm CMI} W_b d_C}$, with the numerically extracted $\gamma_{p,T}^{\rm CMI}$ shown in \cref{fig4}(d) as well. We find that $\gamma_{p,T}^{\rm len}\approx \gamma_{p,T}^{\rm CMI}$, thus confirming our analytical scaling prediction \cref{cmi_predict}.

\section{Classical sampling from noisy 2D random Clifford circuits via MPO-SEBD algorithm}
\label{sec_clif_Sop}

For noisy 2D Clifford circuits, our numerics in \cref{sec_clif_num} and the analytics based on the simplified model in \cref{sec_clif_distri_Sop} indicate that the maximal measurement-induced operator entanglement obeys an area law with scaling $S_{\rm op}^{\rm max}(T,p)\approx T/(\eta p)$. Moreover, stabilizer generators are distributed uniformly in space, and their lengths exhibit exponential decay. Under these conditions, we show below that the MPO-SEBD algorithm \emph{guarantees} classically efficient sampling from noisy 2D random Clifford circuits (i) throughout the entire sub-logarithmic depth regime $T = o(\log N)$ for arbitrarily small constant noise rate $p = \Omega(1)$, and (ii) at constant depths $T = O(1)$ for noise rates $p = \Omega(\log^{-1}N)$, within \emph{polynomial} time.

As outlined in and before \cref{eq:string_ent}, the maximal bond dimension $\chi_{\rm max}$ during each (randomly realized) sampling process can be directly derived from the corresponding maximal measurement-induced operator entanglement within that sample.
Averaged over all realizations of the sampling process, $S_{\rm op}^{\max}\approx T/(\eta p)$ [cf.~\cref{S_model_scale}] and follows an exponentially decaying distribution [cf.~\cref{eq:suppression}], hence the average maximal bond dimension scales as $\langle \chi_{\rm max}\rangle = 2^{T/(\eta p)}$ and follows the same distribution. Following Ref.~\cite{napp2022}, we set a cutoff $\chi_{\rm cutoff}=2^{\Lambda T/(\eta p)}$ with a tunable coefficient $\Lambda$. In the column-by-column sampling [cf.~\cref{fig1}(b)], the MPO-SEBD algorithm alternates between time-step iterations with SVD sweeps that remove only zero singular values. We abort the algorithm whenever a bond dimension exceeds $\chi_{\rm cutoff}$. Thus the algorithm samples exactly from the noisy 2D random Clifford circuit with success rate $1-\varepsilon_{\rm abort}$—equivalently, within total variation distance $\mathrm {TVD} =\varepsilon_{\rm abort}$.

Consider sampling noisy 2D random Clifford circuits of size $L,W\gg1$ and depth $T$. The total number of bond dimension checks performed during the MPO-SEBD evolution is $N_{\rm check}\approx W\times N_b = NT$.
We can upper-bound the total abort probability $\varepsilon_{\rm abort}$ by $N_{\rm check}$ times the largest per-check abort rate, which occurs at $t_{\rm peak}$ and is at most $e^{-\Lambda}$, as implied by the exponentially decaying distribution [cf.~\cref{eq:suppression}]. Hence,
\begin{equation} \label{}
	\varepsilon_{\rm abort} = {\rm TVD}< NTe^{-\Lambda}.
\end{equation}
Putting this together, the MPO-SEBD algorithm samples from a noisy 2D random Clifford circuit with total variation distance bounded by $\mathrm{TVD}<NTe^{-\Lambda}$, with maximal MPO bond dimension $\chi_{\rm cutoff}=2^{\Lambda T/(\eta p)}$. If one aims to achieve a constant accuracy $\mathrm{TVD}<\epsilon$, one can choose $\Lambda=\log(NT/\epsilon)$ and get
$\chi_{\rm cutoff}=\bigl(NT/\epsilon\bigr)^{(\log 2)\cdot T/(\eta p)}$.
In this case, MPO-SEBD runs with polynomial bond dimension, $\chi_{\rm cutoff}={\rm poly}(N)$, for noisy 2D random Clifford circuits of arbitrary constant depth $T=O(1)$ and noise rate $p=\Omega(1)$.
Moreover, if one aims for a system-size–scaled accuracy $\mathrm{TVD}<N\epsilon$ (as in Ref.~\cite{napp2022}), one can choose $\Lambda=\log(T/\epsilon)$ and get
${\chi_{{\rm cutoff}}=2^{T\log(T/\epsilon)/(\eta p)}}$.
In this case, MPO-SEBD runs with polynomial bond dimension for noisy 2D random Clifford circuits of (i) arbitrary sub-logarithmic depths $T=o(\log N)$ and noise rate $p=\Omega(1)$, and (ii) arbitrary constant depths $T=O(1)$ and inverse-logarithmic scaling noise rate $p = \Omega(\log^{-1} N)$. 

We expect this result to yield useful insights into the simulation complexity of noisy, random 2D circuits of sub-logarithmic depth at arbitrarily small, constant noise rates. Moreover, the regime where $p$ scales inversely with $N$ has been investigated in the entanglement dynamics of 1D circuits~\cite{PhysRevB.110.064323}. Our result on the inverse-logarithmic scaling of $p$ provides a nontrivial 2D counterpart, with explicit connection to circuit simulability via the MPO-SEBD algorithm.

Finally, we point out that the mechanism underlying the area-law scaling of MIE based on the stabilizer generators analyzed in \cref{sec_clif_string}, as well as the efficient classical sampling strategy discussed in this section, relies on the Clifford nature of the circuit and does not directly extend to the non-Clifford case. In the next section, we apply MPO-SEBD to  noisy 2D Haar-random circuits to probe the scaling of MIE in that setting. We then discuss in \cref{sec_implic} the qualitative mechanisms that may render the MPO-SEBD algorithm efficient for simulating  noisy 2D non-Clifford circuits.

\section{Demonstration of MPO-SEBD}
\label{sec_haar}
\begin{figure*}
	\centering
	\includegraphics[width=\textwidth]{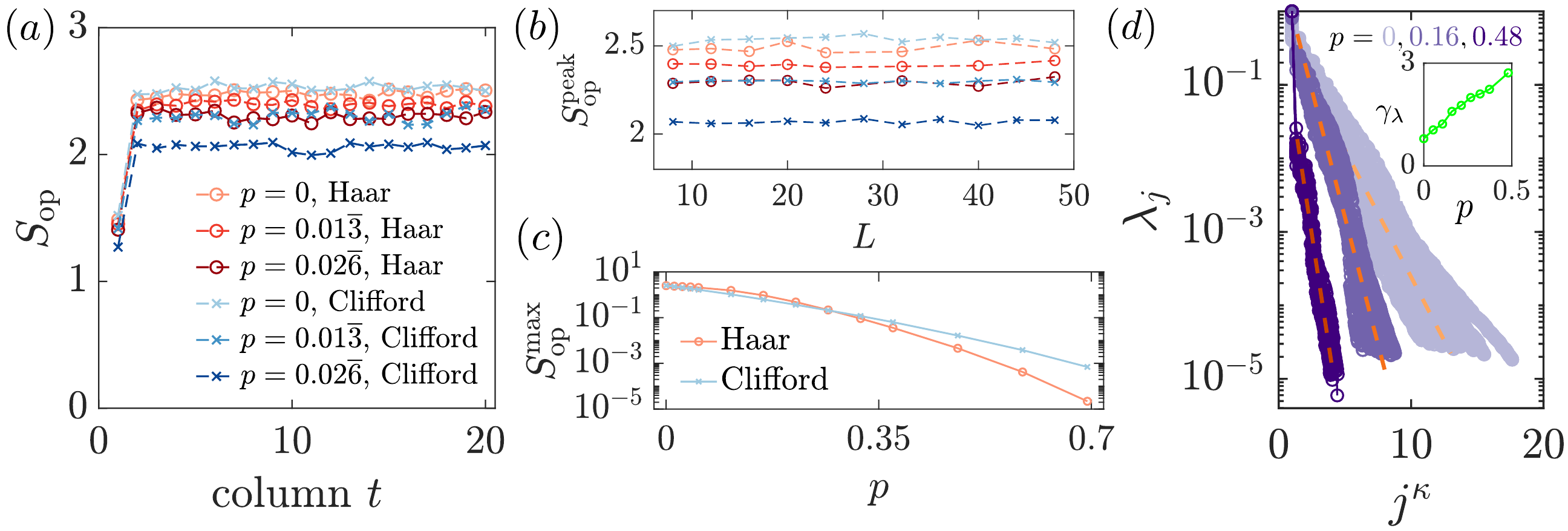}
        \caption{\emph{Comparison of measurement-induced operator entanglement for Haar-random and random Clifford circuits with $T=4$ (within the area-law phase, where $T<T_c$) [cf.~\cref{fig1}(c)].}
\textbf{(a)} Evolution of $S_{\rm op}(t)$ during sampling for both noiseless and noisy circuits. In both cases, $S_{\rm op}(t)$ quickly saturates to a peak value, denoted as $S_{\rm op}^{\rm peak}$ (consistent with the notation used in~\cref{fig2}). $L=12$ here.
\textbf{(b)} Peak value $S_{\rm op}^{\rm peak}$ as a function of system size $L$, exhibiting area-law scaling. The marker styles are identical to those in panel (a). We extract this plateau (as a function of L) value as $S_{\rm op}^{\rm max}$.
\textbf{(c)} Scaling of $S_{\rm op}^{\rm max}$ as a function of the noise rate $p$ (with fixed depth $T=4$).
\textbf{(d)} Behavior of the Schmidt singular values $\lambda_j$ (ordered from largest to smallest) across the half-chain bipartition [cf.~\cref{fig1}(b)] as a function of the index $j$ for the case of length $L=40$. The fitted exponent is $\kappa=0.40$. The dashed line shows the fitted scaling described in \cref{sval_scale}.}
        \label{fig_haar}
\end{figure*}

\subsection{Noisy 2D Haar-random circuits}
\label{haar_num}
Here, we demonstrate the MPO-SEBD algorithm by simulating MIE in the sampling dynamics of noisy 2D Haar-random circuits. We further compare these results with those from random Clifford circuits under the same conditions, providing evidence that the MIE in noisy 2D Haar-random circuits and random Clifford circuits exhibits similar behavior. Note that this is not clear a priori, since Clifford circuits form only a unitary 3-design~\cite{PhysRevA.96.062336}, whereas qualitatively matching entanglement properties may require agreement in higher moments of the Haar-random circuit output distribution.

We consider the same 2D brickwork circuit architecture and employ the MPO-SEBD sampling procedure described in~\cref{sebd_setup}, where the boundary state $\rho_b$ is represented as a MPO~\footnote{In our numerical implementation, we combine the qubits along the width direction into a single qudit site, resulting in a one-dimensional MPO of length $L$. The singular value truncation is performed immediately before the sampling measurement of each column, with the truncation error for each bond \texttt{cutoff} $= 5\times10^{-8}$ in ITensor.jl.}. One important distinction is that while in the Clifford numerics, we simulated the depolarizing noise by sampling from locations, to simulate Haar-random circuit dynamics, we implement the depolarizing channel directly.
For a single-qubit depolarizing channel ${\cal D}_i$ acting on the $i$-th qubit with rate $p$,
\[
{\cal D}_i(\rho)=\sum_{\mu \in \{0,x,y,z \}} K_{\mu,i}\,\rho\,K_{\mu,i}^\dagger,
\]
the Kraus operators are given by
\begin{equation} \label{kraus_dep}
K_{0,i}^{\rm dep}=\sqrt{1-\frac{3p}{4}}\,\mathbb I_i, \qquad
K_{\mu \in\{x,y,z\},i}^{\rm dep}=\sqrt{\frac{p}{4}}\,\sigma_i^\mu,
\end{equation}
where $\mathbb I_i$ and $\sigma_i^{\mu \in\{x,y,z\}}$ denote the identity operator and the Pauli operators acting on the $i$-th qubit, respectively. In the end of the simulation, the measurement-induced operator entanglement [\cref{eq:Sop_def}] is computed by vectorizing the MPO into an MPS.

Since the computational cost of the MPO-SEBD algorithm grows exponentially with the circuit depth $T$ [cf.~\cref{sebd_setup}], our numerical simulations for large systems are restricted to circuits of depth $T=4$. Note that a 2D circuit of depth $T=4$ is already worst-case hard to sample~\cite{Bouland2019a}. For Haar-random circuits, the MIE is also expected to undergo a finite-depth transition from area- to volume-law scaling at some critical depth $T_c^{\rm haar} > 3$~\cite{napp2022}. It has not been explicitly demonstrated whether depth $T = 4$ lies on the area-law or volume-law side of this transition. However, our numerics based on random Clifford circuits suggest that for $T=4$, the MIE likely exhibits an area law even in the absence of noise. Indeed, the results in this section indicate that Haar-random circuits and random Clifford circuits display the same behavior at $T = 4$.

The typical evolution of $S_{\rm op}(t)$ for the boundary state during the sampling process [cf.~\cref{fig1}(b)] for both Haar-random circuits and random Clifford circuits is shown in \cref{fig_haar}(a). In both noiseless and noisy cases, $S_{\rm op}(t)$ rapidly saturates to a steady-state value, denoted as $S_{\rm op}^{\rm peak}$. Note that $S_{\rm op}(t)$ does not exhibit the rise-and-fall behavior seen in \cref{fig2}(a), because $T=4$ lies on the area-law side of the phase diagram even in the absence of noise. We also observe that introducing noise monotonically reduces $S_{\rm op}^{\rm peak}$, and that the results for Haar-random circuits and random Clifford circuits under the same noise rate $p$ show good qualitative agreement. 

Using the MPO-SEBD algorithm, we perform simulations for lattice lengths up to $L=48$ and plot the extracted $S_{\rm op}^{\rm peak}$ as a function of $L$ in~\cref{fig_haar}(b), which clearly exhibits area-law scaling for both noiseless and noisy cases. Moreover, we find that the peak values of the operator entanglement for individual random realizations follow a Gaussian-like distribution centered around their average value $S_{\rm op}^{\rm peak}$. These results demonstrate that 2D Haar-random circuits of depth $T=4$ obey area-law scaling of MIE, suggesting that they may be efficiently simulated classically using the SEBD algorithm~\cite{napp2022} in the noiseless case and the MPO-SEBD algorithm introduced here in the noisy case.

We further present the scaling of the asymptotic (in $L$) $S_{\rm op}^{\rm max}$ with respect to the noise rate $p$ in \cref{fig_haar}(c), again demonstrating that the Haar-random and random Clifford cases show good qualitative agreement. Note that here, $S_\mathrm{op}^\mathrm{max}$ decays almost exponentially with $p$ rather than as $1/p$ as we observed for random Clifford circuits of depth $T>T_c$.
We attribute this radical change in behavior to the fact that at $T=4<T_c$, there is no volume-law phase at zero noise rate [see \cref{fig1}(c)].

We also analyzed the MPO Schmidt singular values $\lambda_j$ across the half-chain bipartition of the boundary state $\rho_b$ in \cref{fig_haar}(d)~\footnote{Here we focus exclusively on the MPO Schmidt singular values for Haar-random circuits. In contrast, for Clifford circuits all Schmidt singular values are identical, $\lambda_j=\lambda$ for all $j$, and are directly determined by the operator entanglement via $S_{\rm op}=-\log \lambda^2$.}. Since the system exhibits an area law in both the noiseless and noisy cases at $T = 4$, we observe a similar behavior of $\lambda_j$ (ordered from largest to smallest) in these two cases, which approximately follow
\begin{equation} \label{sval_scale}
	\lambda_j\sim \exp(-\gamma_{ \lambda}(p)\cdot  j^{\kappa}),
\end{equation}
where the exponent $\kappa=0.40$ and the mean coefficient $\gamma_\lambda(p)$ are obtained from numerical fitting. The coefficient $\gamma_\lambda(p)$ increases monotonically with $p$, again demonstrating that $S_{\rm op}$ decreases monotonically as noise increases. This rapid stretched-exponential decay of the MPO singular values provides complementary evidence for the area-law scaling of $S_{\rm op}^{\rm max}$.

The above results demonstrate the MPO-SEBD algorithm for 2D Haar-random circuits with system size up to $L=48$ and depth $T=4$, showing that noise suppresses MIE, which is expected to reduce the simulation complexity of such circuits. Furthermore, across all numerical examples presented, we observe good qualitative agreement between the behavior of Haar-random circuits with deterministic depolarizing noise and that of random Clifford circuits with depolarizing noise modeled in a Monte Carlo manner. This consistency indicates that the MIE properties studied for noisy 2D random Clifford circuits can provide valuable insights into the MIE behavior of Haar-random circuits under similar conditions. We note, however, that this conclusion does not directly extend to the regime $T > T_c$, which we did not explore numerically due to computational constraints. A discussion of this regime of Haar-random circuits is provided in \cref{sec_implic}.

\subsection{Noisy 2D MBQC-type circuits}
 \label{mbqc_sec}

\begin{figure}[b!]
	\centering
	\includegraphics[width=0.45\textwidth]{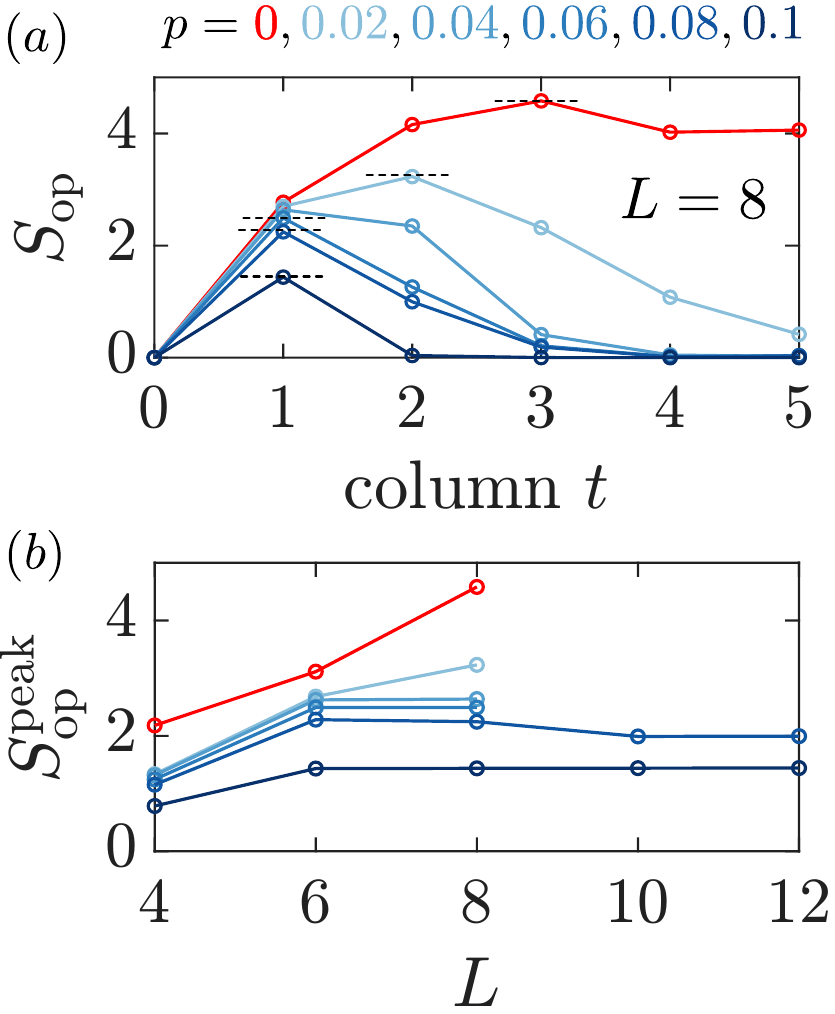}
        \caption{\textit{Results for 2D MBQC-type circuits}. (a) Evolution of $S_{\rm op}(t)$ during the sampling process for MBQC-type circuits under various noise rates $p$, with $L=8$. The dashed lines indicate the corresponding peak values $S_{\rm op}^{\rm peak}$.
\textbf{(b)} Peak value $S_{\rm op}^{\rm peak}$ as a function of system size $L$. The marker styles are identical to those used in panel (a).}
        \label{cls_fig}
\end{figure}

To further demonstrate that noise drives the MIE from a volume-law phase to an area-law phase in non-Clifford circuits, we apply the MPO-SEBD algorithm to a 2D MBQC-type circuit of depth $T=4$, which has been conjectured to be hard to simulate classically~\cite{Bermejo-Vega2018}. We consider a 2D lattice of size $N=L\times W$ [see \cref{fig1}(a,b)], with qubits initialized in the product state
\begin{equation} \label{}
\left|\psi_\beta\right\rangle\propto    \bigotimes_{i=1}^N\left( |0\rangle+\mathrm{e}^{\mathrm{i} \beta_i}|1\rangle\right),
\end{equation}
where the phase $\beta_i$ at site $i$ is chosen independently and uniformly at random from $\{0,\pi/4\}$. We then apply $T=4$ layers of commuting CZ gates according to the schedule shown in \cref{fig1}(a), thereby entangling the lattice, followed by measurements in the $X$ basis. This circuit corresponds to Architecture I introduced in Ref.~\cite{Bermejo-Vega2018}, where sampling from its output distribution was conjectured to be classically hard. This, in turn, suggests that the measurement-induced operator entanglement $S_{\rm op}$ exhibits volume-law scaling.

We apply the MPO-SEBD algorithm to simulate the sampling process of the above MBQC-type circuit in the presence of single-qubit depolarizing noise. The MPO-SEBD iteration procedure is the same as that used for the depth-$T=4$ 2D Haar-random circuit in~\cref{haar_num}, since the two circuits share the same layout. Note that, because the MBQC-type circuit generates a large amount of MIE, our simulations are restricted to relatively small system sizes. To mitigate finite-size effects, we characterize the MIE by the measurement-induced operator entanglement $S_{\rm op}$ of the post-measurement state, maximized over all bipartitions.

The evolution of $S_{\rm op}(t)$ for various depolarizing noise rates $p$ and fixed length $L=8$ is shown in \cref{cls_fig}(a). For the noiseless circuit ($p=0$), the $S_{\rm op}$ grows during the sampling process and saturated at a considerable value. In the noisy cases, by contrast, $S_{\rm op}$ exhibits a rise-and-fall profile, indicating that the MIE growth is suppressed by the accumulation of noise during the sampling evolution. This behavior is in good qualitative agreement with that of noisy random Clifford circuits in the regime $T>T_c$ [cf.~\cref{fig2}(a)].

We extract the maximal value of $S_{\rm op}$ during the evolution, denoted by $S_{\rm op}^{\rm peak}$, and show its scaling with system size $L$ for various fixed noise rates $p$ in \cref{cls_fig}(b). In the noiseless case, $S_{\rm op}^{\rm peak}$ grows approximately linearly with $L$, consistent with the expected volume-law scaling of the MIE in the absence of noise~\cite{Bermejo-Vega2018}. Upon introducing noise, $S_{\rm op}^{\rm peak}$ is systematically suppressed, and for sufficiently large noise rates we observe a plateau-like saturation of $S_{\rm op}^{\rm peak}$, suggesting an area-law scaling in this regime. These results therefore provide numerical evidence that depolarizing noise can destroy the volume-law scaling of MIE in 2D non-Clifford circuits.

\section{Discussions}
\label{sec_implic}

\emph{Classical algorithm based on exponentially decaying CMI.---}Although our analysis mainly focuses on the measurement-induced operator entanglement and the MPO-SEBD algorithm, it has implications for other classical algorithms as well.
We observed an exponential decay of the conditional mutual information in the boundary state, but one could conjecture that this holds for any choice of $A,B,C$ in the full lattice where $A$ and $B$ are separated by some buffer that includes $C$.
Such an `approximate markov' property would lead to provable classical simulability via `patching'-style algorithms~\cite{napp2022,PRXQuantum.6.010356}, which are motivated by techniques from Gibbs sampling~\cite{Brandao2019}, and it can lead to representability by classical neural networks as well~\cite{yang2024classicalneuralnetworksrepresent}.

\emph{Extension to Haar random circuits of depth $T>T_c$.---}Even though the numerics and results presented here concern Clifford circuits, understanding the complexity of sampling from noisy random circuits containing general gates is a key motivation for this work.
The MPO-SEBD algorithm directly extends to that setting. We expect the same behavior of the maximal measurement-induced operator entanglement in this setting, but verifying this numerically becomes substantially more challenging beyond the depth $T = 4$ case studied in \cref{sec_haar}.

Analytical support for this expectation can be obtained from a statistical mechanics mapping of Haar-random circuits. Following the procedure of Ref.~\cite{napp2022}, one can view the SEBD algorithm as exhibiting dynamics similar to a 1D Haar random circuit with interleaved weak measurements. In Ref.~\cite{napp2022}, such a mapping was shown to exhibit a transition between an ordered phase and a disordered phase as a function of circuit depth. This corresponds to a transition from an area-law to a volume-law in certain entropic quantities that are heuristically expected to reflect the efficiency of the SEBD algorithm—an area (respectively volume) law corresponding to efficient (respectively inefficient) runtime. 
In our case, the key difference is the presence of depolarizing noise in addition to the measurements. This noise acts as a symmetry-breaking field, suppressing the ordered phase and preventing a sharp transition. These observations are consistent with the operator entanglement entropy in the quasi-1D noisy circuit being in an area-law phase (see, e.g., Ref.~\cite{li2023entanglement} for similar arguments in 1D noisy circuits). Similar symmetry-breaking effects due to noise have been observed in other settings, including 1D monitored circuits with bulk noise~\cite{Dias_noise_symmetry_breaking, Li_dissipative_2023,qian_protect_PRL_2025}, as well as 1D noisy quantum circuits without measurements~\cite{li2023entanglement,Niroula_mitigation_2025}.

Note that even if the measurement-induced operator entanglement, $S_{\rm op}^{\rm max}$, obeys an area-law scaling, extending the MPO-SEBD algorithm to generic non-Clifford circuits (e.g., Haar-random circuits) will require truncating singular values. However, truncation guarantees are typically formulated in the $L_2$ norm rather than the $L_1$ norm, so an area law for $S_{\rm op}^{\rm max}$ alone does not establish efficient sampling in total variation distance.  Regarding this issue, Ref.~\cite{wei2026noise} provides evidence that, in the setting of 1D Haar-random circuits and 1D Lindbladian dynamics with either depolarizing noise or amplitude-damping noise, MPO truncation errors contract exponentially under the noisy dynamics. This leads to improved empirical bounds on the $L_1$ truncation error and suggests that MPO-based simulation may serve as an accurate algorithm in $L_1$ error for these settings. We expect a similar noise-induced error contraction to occur in simulations of  noisy 2D non-Clifford circuits using the MPO-SEBD algorithm. Another possible direction could be to consider the renormalization-group–style truncation introduced in Ref.~\cite{PRXQuantum.1.010304}, which provides $L_1$-norm error control when \textit{entanglement of purification} of the boundary states $\rho_t$ during the sampling evolution obeys an area law.

\emph{Concurrent works.---}While finalizing this manuscript, we became aware of Refs~\cite{lee2025classical,zhang2025classicallysamplingnoisyquantum}, which extend the patching algorithm of Ref.~\cite{napp2022} to sample from noisy random circuits under the assumption of exponentially decaying CMI. In particular, Ref.~\cite{lee2025classical} provides numerical evidence for exponential CMI decay in 1D noisy Haar-random circuits and in noisy 2D Clifford circuits with both unital and non-unital noise. Ref.~\cite{zhang2025classicallysamplingnoisyquantum} presents evidence of exponential CMI decay for noisy 2D random circuits using Clifford numerics and an analytical mapping to a statistical-mechanical model for noisy 2D Haar-random circuits at large qudit dimension, and further proves exponential CMI decay for arbitrary circuits when the noise rate exceeds a threshold $p_c>0$.

Their numerical results on CMI decay for noisy 2D random Clifford circuits agree with ours. Notably, for noisy 2D random Clifford circuits, exponential CMI decay implies that the patching algorithm~\cite{napp2022,PRXQuantum.6.010356,lee2025classical,zhang2025classicallysamplingnoisyquantum,nelson2025limitationsnoisygeometricallylocal} runs in polynomial time for depths $T=O(\sqrt{\log N})$ and in \emph{quasi-polynomial} time beyond that. By contrast, an area-law scaling of the maximal operator entanglement $S_{\rm op}^{\rm max}$ implies that the MPO-SEBD algorithm scales \emph{polynomially} throughout the entire sub-logarithmic depths $T=o(\log N)$ for noisy 2D Clifford circuits under \emph{any} constant noise. It is therefore an interesting open question whether MPO-SEBD can be leveraged into a provably efficient classical algorithm for sampling from general (non-Clifford) noisy 2D circuits of sub-logarithmic depths at any constant noise rate.

\section*{Acknowledgements}
We thank Benjamin Remez, Ignacio Cirac, Jiyao Chen, Jens Eisert, Ruijing Guo, Tsung-Cheng Lu, Yifan Zhang, Ziyang Zhang for insightful discussions. ZYW, JR, and AVG acknowledge support from the U.S.~Department of Energy, Office of Science, Accelerated Research in Quantum Computing, Fundamental Algorithmic Research toward Quantum Utility (FAR-Qu). ZYW and AVG were also supported in part by NSF QLCI (award No.~OMA-2120757), NQVL:QSTD:Pilot:FTL, NSF STAQ program, DoE ASCR Quantum Testbed Pathfinder program (awards No.~DE-SC0019040 and No.~DE-SC0024220), ONR MURI,  AFOSR MURI,  DARPA SAVaNT ADVENT, and ARL (W911NF-24-2-0107). ZYW and AVG also acknowledges support from the U.S.~Department of Energy, Office of Science, National Quantum Information Science Research Centers, Quantum Systems Accelerator (QSA).
M.J.G, J.N., and J.R. acknowledge support from the NSF QLCI award OMA2120757. This work was performed in
part at the Kavli Institute for Theoretical Physics (KITP), which is supported by grant NSF PHY-2309135.
JN is supported by the National Science Foundation Graduate Research Fellowship Program under Grant No. DGE 2236417. DM acknowledges financial support by the Novo Nordisk Foundation under grant numbers NNF22OC0071934 and NNF20OC0059939. E.C. acknowledges the Munich Quantum Valley, which is supported by the Bavarian state government with funds from the Hightech Agenda Bayern Plus. E.C. further acknowledges funding from the German Federal Ministry of Education and Research (BMBF) through EQUAHUMO (Grant No. 13N16066) within the funding program Quantum Technologies—From Basic Research to Market. 
We acknowledge the use of AI tools for English language refinement. The numerical calculations were
performed using the QuantumClifford.jl~\cite{QClif} and ITensor.jl~\cite{itensor}.

\appendix

\newpage
\bibliography{library.bib, biblio_esther.bib}

\begin{thebibliography}{60}
\bibitem [{vft()}]{vftn_fn}%
  \BibitemOpen
  \href@noop {} {\bibinfo  {journal} {The exact value of $W_b$ is set by the
  brick-wall geometry and the past light cone of the boundary state shown in
  Fig.~1(a,b), and varies with the sampling column $t$ during the sampling
  process. For example, when the circuit depth is $T=4n$, the boundary state
  before sampling $t$-th column of qubits with even $t$, one has $W_b=2n-1$.}\
  }\BibitemShut {NoStop}%
\end{thebibliography}
\end{document}